\documentclass[final]{siamltex}

\usepackage{amsmath,amssymb,graphicx}
\usepackage[dvips]{epsfig}
\usepackage{showkeys}

\def \R{\mathbb R}

\newcommand{\lapprox} {\, \lower3pt\hbox{$\sim$}\llap{\raise2pt\hbox{$<$}}\,}
\newcommand{\gapprox} {\, \lower3pt\hbox{$\sim$}\llap{\raise2pt\hbox{$>$}}\,}

\newcommand{\x}{\textbf{x}}
\newcommand{\y}{\textbf{y}}
\newcommand{\z}{\textbf{z}}
\newcommand{\kr}{{\bf{k}}^r}
\newcommand{\kf}{{\bf{k}}^f}

\newcommand{\xik}{\boldsymbol{\xi}}
\newcommand{\zero}{\textbf{0}}

\newtheorem{remark}{Remark}[section]

\title{The process of data formation for the {\em{Spectrometer/Telescope for Imaging X-rays (STIX)}} in {\em{Solar Orbiter}}}

\author{
S.~Giordano\thanks{Dipartimento di Matematica, Universit\`a di Genova, via Dodecaneso 35, 16146 Genova, Italy ({\tt giordano@dima.unige.it}).} \and
N.~Pinamonti\thanks{Dipartimento di Matematica, Universit\`a di Genova, via Dodecaneso 35, 16146 Genova, Italy and INFN, Sezione di Genova, via Dodecaneso 33, 16146 Genova, Italy  ({\tt pinamonti@dima.unige.it}).} \and
M.~Piana\thanks{Dipartimento di Matematica, Universit\`a di Genova and CNR - SPIN, Genova, via Dodecaneso 35, 16146 Genova, Italy ({\tt piana@dima.unige.it}).} \and
A.~M.~Massone\thanks{CNR - SPIN, Genova, via Dodecaneso 33, I-16146 Genova, Italy ({\tt annamaria.massone@cnr.it}).} }

\begin{document}

\maketitle

\begin{abstract}
The {\em{Spectrometer/Telescope for Imaging X-rays (STIX)}} is a hard X-ray imaging spectroscopy device to be mounted in the {\em{Solar Orbiter}} cluster with the aim of providing images and spectra of solar flaring regions at different photon energies in the range from a few keV to around $150$ keV. The imaging modality of this telescope is based on the Moir\'e pattern concept and utilizes $30$ sub-collimators, each one containing a pair of co-axial grids. This paper applies Fourier analysis to provide the first rigorous description of the data formation process in {\em{STIX}}. Specifically, we show that, under first harmonic approximation, the integrated counts measured by {\em{STIX}} sub-collimators can be interpreted as specific spatial Fourier components of the incoming photon flux, named {\em{visibilities}}. Fourier analysis also allows the quantitative assessment of the reliability of such interpretation. The description of {\em{STIX}} data in terms of visibilities has a notable impact on the image reconstruction process, since it fosters the application of Fourier-based imaging algorithms. 
\end{abstract}


\pagestyle{myheadings}

\thispagestyle{plain}

\section{Introduction}
The {\em{Spectrometer/Telescope for Imaging X-rays (STIX)}} \cite{beetal12} is a hard X-ray instrument mounted as part of the {\em{Solar Orbiter}} cluster which will be launched by the {\em{European Space Agency (ESA)}} in 2018.  The main scientific goal of the {\em{STIX}} mission is to measure hard X-ray photons emitted during solar flares in order to determine the intensity, spectrum, timing and location of accelerated electrons near the Sun. 
This imager and spectrometer is formed by $30$ detectors recording $X-$ray photons in the range $4-150$ keV (two more detectors are added, which play the role of a coarse flare locator and a background monitor). On each detector, the incident flux is modulated by means of a sub-collimator formed by two distant grids with slightly different pitches and slightly different orientations. The effect of this grid configuration is to create the superposition of two spatial modulations, named {\em{Moir\'e pattern}} \cite{oswazw64}. The recording process on the detector associated to each Moir\'e pattern provides a spatial Fourier component of the incoming flux, named {\em{visibility}}. Therefore {\em{STIX}} recording hardware allows sampling the spatial frequency domain, named $(u,v)$ plane, in $30$ different points. 

The use of visibilities in astronomical imaging is traditionally related to radio interferometry \cite{thetal01}. However more recently this modality has been utilized in the case of hard X-ray telescopes like Yohkoh {\em{Hard X-ray Telescope (HXT)}} \cite{koetal91} and the {\em{Reuven Ramaty High Energy Solar Spectroscopic Imager (RHESSI)}} \cite{lietal02,huetal02}. The general goal of the present paper is to study the model for the formation of the visibility signal in {\em{STIX}}. In particular, we provide the first rigorous description of the process that, starting from the incoming photon flux, leads to the identification of the photon counts recorded by the detectors with the visibilities. Such a description is based on the numerical approximation according to which the transmission function of each sub-collimator grid is represented by a sine function. Therefore, the second aim of the paper is to compute the approximation error introduced by this first harmonic assumption and to determine to what extent the interpretation of {\em{STIX}} data in terms of visibilities is quantitatively reliable. 

The plan of the paper is as follows. Section 2 describes the hardware characteristics of the instrument. Section 3 contains the computation of the {\em{STIX}} visibilities. Section 4 provides an analysis of the interpretation error induced by identifying the counts recorded by the detectors with the photon visibilities. Our comments and conclusions are offered in Section 5. 

\section{Description of the instrument}
\begin{figure}
\begin{center}
\includegraphics[width=9.5cm]{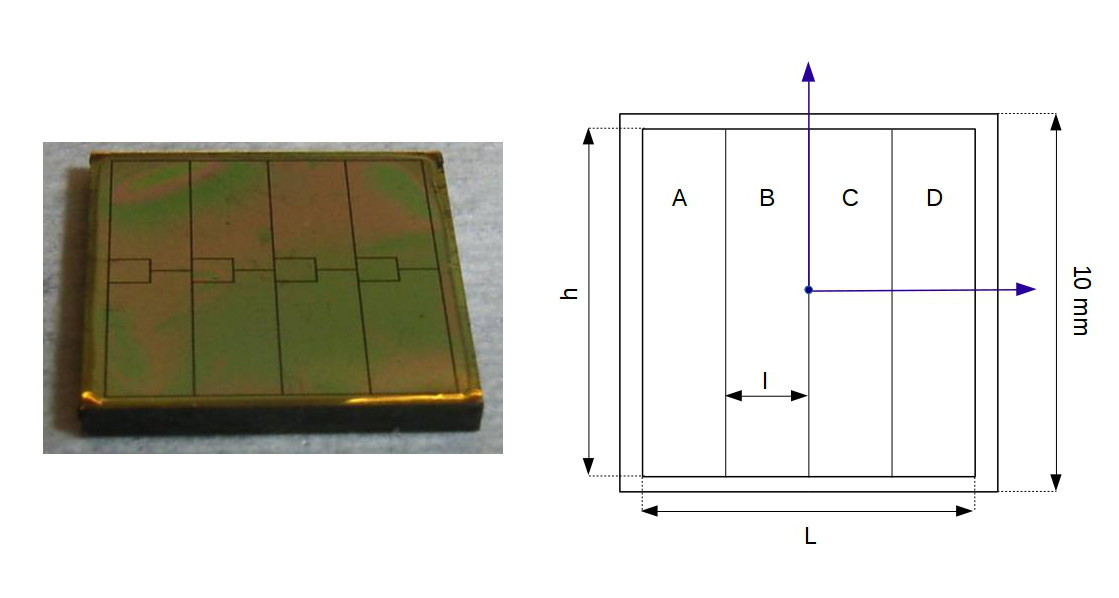}
\end{center}
\caption{The {\em{STIX}} detectors. Left panel: a snapshot of one of the {\em{STIX}} detectors. Right panel: schematic representation of the four pixels for each detector and notations.}
\label{fig:detector}
\end{figure}

{\em{STIX}} conveys X-rays from the Sun through $30$ pairs of tungsten grids mounted at the extremity of an aluminum tube (two more detectors, labelled with numbers $9$ and $10$, do not produce Moir\'e patterns and are used as flare locator and background monitor, respectively). The modulation pattern is recorded by a pixelized Cadmium-Telluride detector \cite{poetal13} made of four rectangular pieces and mounted behind each grid pair. We now introduce the notations and describe the physical and geometric parameters  that will be used to compute the radiation pattern modulated by the grid pairs and the signal recorded by the detectors.
\begin{description}
\item{\bf{Global parameters:}} these parameters describe the {\em{STIX}} large scale structure and its position with respect to the Sun:
\begin{itemize}
\item $S$ is the distance of {\em{Solar Orbiter}} orbit from the Sun. This distance will significantly change during the mission. For sake of simplicity and accordingly with the value currently used in the {\em{STIX}} simulation software, in this paper we will adopt $S=1$ astronomic unit, i.e. $S \simeq 1.5 \times 10^8$ km.
\item $L_1= 55$ cm is the separation distance between the front and the rear grids, i.e. the length of the aluminum tube.
\item $L_2=4.7$ cm is the separation distance between the rear grids and the detectors.
\end{itemize}
\item{\bf{Detectors and pixels:}} each detector (see Figure \ref{fig:detector}) is a square with side $10$ mm; inside this square the region sensitive to photons is a rectangle of dimensions $L \times h$ where $L=8.8$ mm and $h=9.2$ mm. In a standard acquisition mode, this rectangle is divided into four identical pixels with dimensions $l \times h$, where $l=2.2$ mm. The four pixels are denoted with $A$, $B$, $C$, and $D$. 
\item{\bf{Grids:}} a pair of grids made of a front grid and a rear grid is associated to each detector, forming a so called sub-collimator. The front and rear grids are made by equally spaced slits and slats and therefore are characterized by constant pitches $p^f$, $p^r$ and orientations $\alpha^f$, $\alpha^r$, respectively. These latter information are combined into the wave vectors
\begin{equation}\label{wavevector}
{\bf{k}}^f =(k^f_1,k^f_2)= \left(\frac{\cos \alpha^f}{p^f},\frac{\sin\alpha^f}{p^f}\right)~~~~~~{\bf{k}}^r=(k^r_1,k^r_2)= \left(\frac{\cos \alpha^r}{p^r},\frac{\sin\alpha^r}{p^r}\right)~.
\end{equation}
\item{\bf{Pitches and orientations:}} {\em{STIX}} sub-collimators present $10$ different pitches geometrically increasing from $0.038$ mm to $0.953$ mm with ratio $1.43$, and $9$ uniformly increasing orientation angles from $10^{\circ}$ to $170^{\circ}$ with step $20^{\circ}$. Pitches and orientations for the front and rear grids are not perfectly identical, and therefore ${\bf{k}}^f \neq {\bf{k}}^r$, but they are built in such a way that
\begin{equation}\label{difference-construction}
{\bf{k}}^f - {\bf{k}}^r = (\pm 1/L,0).
\end{equation}
In the following all computations will be performed by taking the plus sign in equation (\ref{difference-construction}), but the approach is analogous in the case of the minus sign.
\end{description}

\begin{figure}
\begin{center}
\includegraphics[width=7.cm]{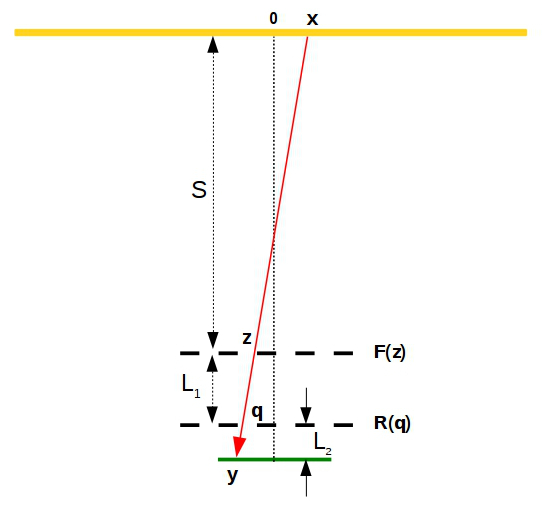}
\end{center}
\caption{Schematic representation of the signal transmission process through one of the {\em{STIX}} sub-collimators. The red line is the path travelled by a photon emitted from the point ${\bf{x}}$ on the Sun (yellow surface) and reaching the point ${\bf{y}}$ on the detector (in green).
}
\label{fig:STIX-diagram}
\end{figure}

\section{Signal formation}
In order to describe the signal formation process in {\em{STIX}} we introduce four parallel two-dimensional coordinate systems determined, respectively, by the detector surface, the two grids and the Sun surface. Then we consider the following simple scenario: from the point ${\bf{x}}=(x_1,x_2)$ on the Sun disk, the photon flux $\phi({\bf{x}})$ reaches, in a time interval $\Delta t$, the front grid at ${\bf{z}}=(z_1,z_2)$, the rear grid at ${\bf{q}}=(q_1,q_2)$, and finally the detector surface at ${\bf{y}}=(y_1,y_2)$ (see Figure \ref{fig:STIX-diagram}).
Standard relations between similar triangles lead to 
\begin{equation}\label{zxy}
\z \quad = \quad \y - \frac{(\x + \y)(L_1+L_2)}{S+L_1+L_2} \quad \simeq \quad \y - \x\frac{L_1+L_2}{S}
\end{equation}
and
\begin{equation}\label{txy}
\textbf{q} \quad = \quad {\bf{y}} - \frac{L_2(\x + \y)}{S+L_1+L_2} \quad \simeq \quad \y - \x\frac{L_2}{S}~,
\end{equation}

\noindent which implies that ${\bf{z}}={\bf{z}}({\bf{x}},{\bf{y}})$ and ${\bf{q}}={\bf{q}}({\bf{x}},{\bf{y}})$. We point out that the approximation in these two equations is reliable because it neglects  terms of the order of magnitude of centimeters with respect to terms of the order of magnitude of millions of chilometers.

\begin{figure}
\begin{center}
\begin{tabular}{cc}
\hspace{-0.5cm}
\includegraphics[width=6.5cm]{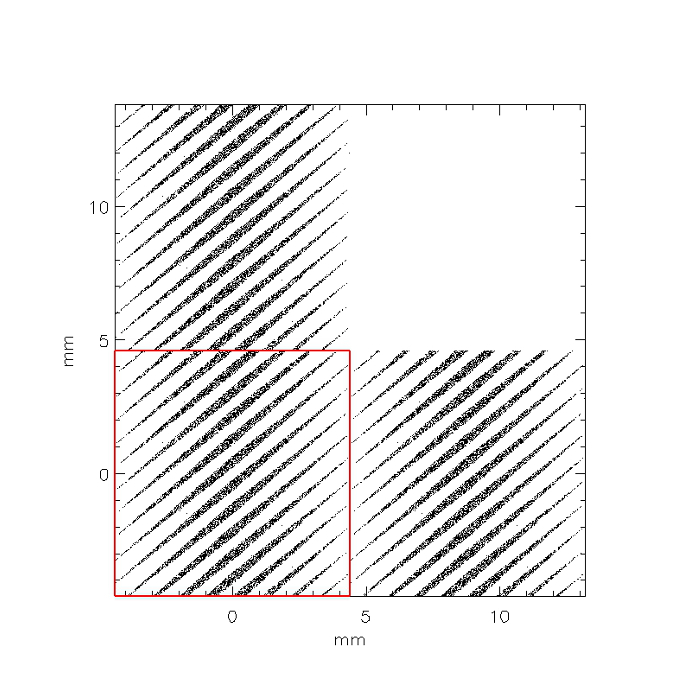} &
\hspace{-1.5cm}
\includegraphics[width=6.5cm]{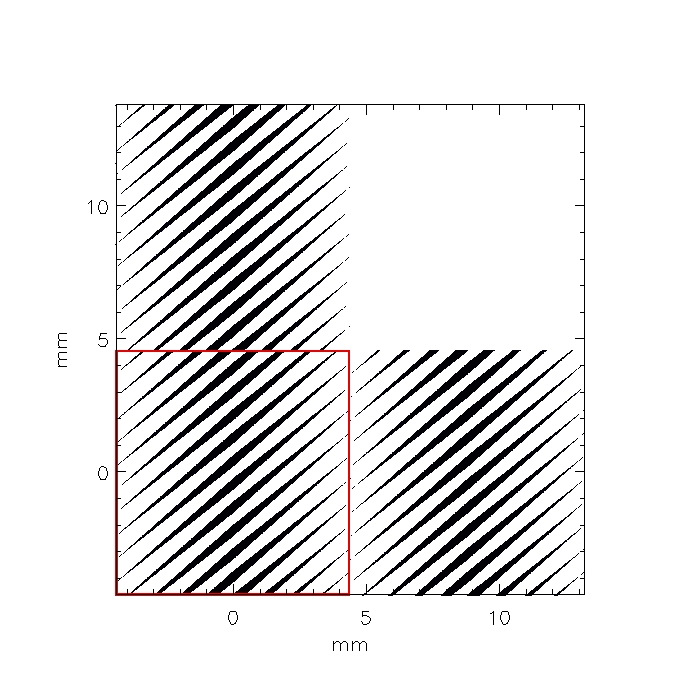}
\end{tabular}
\end{center}
\caption{Transmission function of sub-collimator $32$ for ${\bf{x=0}}$. Left panel: simulation obtained by means of Monte Carlo computation. Right panel: transmission function obtained by using formulas (\ref{bo-1})-(\ref{product}) with $100$ harmonics. For each panel, we reported the transmission function (highlighted in the bottom left corner) and two replicas that allow checking the periodicity along both $y_1$ and $y_2$ directions.}
\label{fig:pattern}
\end{figure}

The values $F({\bf{z}})$ and $R({\bf{q}})$ of the transmission functions through the front and rear grids at points ${\bf{z}}$ and ${\bf{q}}$ respectively, can be modeled as two-dimensional step functions and therefore computed by means of a Fourier series as \cite{crwy92}
\begin{equation}\label{comput-4}
F({\bf{z}}) = \frac{1}{2} + \frac{2}{\pi}\sum_{m=1,3}^{\infty}\frac{1}{m} \sin\left(2\pi m \kf \cdot ({\bf{z}} + {\bf{t}}) \right)
\end{equation}
and
\begin{equation}\label{comput-5}
R({\bf{q}}) = \frac{1}{2} + \frac{2}{\pi}\sum_{j=1,3}^{\infty}\frac{1}{j} \sin\left(2\pi j \kr \cdot ({\bf{q}} + {\bf{t}}) \right)~,
\end{equation}
where the constant vector ${\mathbf{t}}=(0,t)$ accounts for a possible $y_2$-translation of the grid pair with respect to the origin (in the current version of the {\em{STIX}} simulation software $t = 1/(4 k_2^f) =  1/(4 k_2^r)$).
Exploiting approximations (\ref{zxy}) and (\ref{txy}), equations (\ref{comput-4}) and (\ref{comput-5}) can be represented in terms of ${\bf{x}}$ and ${\bf{y}}$ as
\begin{equation}\label{bo-1}
F(\x,\y) = \frac{1}{2} + \frac{2}{\pi}\sum_{m=1,3}^{\infty}\frac{1}{m} \sin\left(2\pi m \kf \cdot \left(\y - \x \frac{L_1 + L_2}{S} + {\mathbf{t}} \right) \right)
\end{equation}
and
\begin{equation}\label{bo-2}
R(\x,\y) = \frac{1}{2} + \frac{2}{\pi}\sum_{j=1,3}^{\infty}\frac{1}{j} \sin\left(2\pi j \kr \cdot \left(\y - \x \frac{L_2}{S} + {\mathbf{t}} \right) \right)~.
\end{equation}
The global transmission function for each {\em{STIX}} sub-collimator is given by
\begin{equation}\label{product}
T(\x,\y) = F(\x,\y)R(\x,\y)~.
\end{equation}
Figure \ref{fig:pattern} contains a representation of this function for $\x=0$ in the case of sub-collimator 32, computed by means of both formulas (\ref{bo-1})-(\ref{product}), when the sums in (\ref{bo-1}),(\ref{bo-2}) are truncated at $100$ series components, and a Monte Carlo computation provided by {\em{STIX}} software for $100000$ photons per cm$^2$. In each panel, the pattern highlighted in red at the left, bottom corner is the actual Moir\'e pattern associated to the sub-collimator, while the replicas on the right and on the top allow one to check the periodicity of such a pattern along both  the $y_1$ and the $y_2$ directions.  The figure clearly shows that this specific sub-collimator (sub-collimator $32$) exactly samples one full period in neither directions. We have investigated this behavior for all sub-collimators and found that $10$ sub-collimators over $30$ (sub-collimators $2$, $4$, $5$, $8$, $21$, $23$, $24$, $26$, $31$, and $32$) have periodicity neither along $y_1$ nor along $y_2$; $9$ sub-collimators present periodicity along $y_1$ but not  along $y_2$ (sub-collimators $3$, $6$, $7$, $11$, $16$, $18$, $20$, $22$, and $28$); $8$ sub-collimator present periodicity along $y_2$ but not along $y_1$ (sub-collimators $1$, $12$, $14$, $15$, $17$, $27$, $29$, and $30$); finally, the remaining $3$ sub-collimators (sub-collimators $13$, $19$, and $25$) have periodicity in both directions.

%
%
%
%

We want now to determine the number of counts recorded by each pixel when reached by the flux modulated by the two grids. To this aim we first introduce the following definition, which is systematically used in radio interferometry and hard X-ray imaging:
\begin{definition}\label{def-visibility}
Given the scalar field $\phi \in L^1(\R^2)$ representing the photon flux incoming from the Sun to the telescope, the Fourier transform 
\begin{equation}\label{eq:visibilities}
V(\xik) = \int  \phi({\bf{x}}) \exp(i 2 \pi \xik \cdot {\bf{x}}) d{\bf{x}}~,
\end{equation}
of $\phi({\bf{x}})$ computed at $\xik \in \R^2$ is named the visibility associated to $\phi({\bf{x}})$ and computed at the point $\xik=(u,v)$ in the $(u,v)$ plane of the spatial frequencies conjugated to ${\bf{x}}$.
\end{definition}

Then the following theorem holds true:
\begin{theorem}\label{theo-counts}
Under the approximation (\ref{zxy}) and (\ref{txy}) and if just the first harmonic component is kept in equations (\ref{bo-1}) and (\ref{bo-2}), the number of counts recorded by the $n$-th pixel is given by
\begin{equation}\label{counts-3}
\begin{split}
{\cal{C}}_n = & M_0V({\bf{0}}) + H_{n}^{(1,1)}V \left(-\kf\frac{L_1+L_2}{S}\right) 
                              + H_n^{(1,2)}V \left(\kf\frac{L_1+L_2}{S}\right) + \\
                          & + H_n^{(2,1)}V \left(-\kr\frac{L_2}{S}\right) + H_n^{(2,2)}V \left(\kr\frac{L_2}{S}\right) + \\
                          & + H_n^{(3,1)}V \left(-\kf\frac{L_1+L_2}{S}-\kr\frac{L_2}{S}\right) + H_n^{(3,2)}V
                              \left(\kf\frac{L_1+L_2}{S}+\kr\frac{L_2}{S}\right) + \\
                          & + M_1\exp\left(i\frac{\pi}{2}n\right)\exp\left(i\frac{\pi}{4}\right)V \left(-\kf\frac{L_1+L_2}{S}+\kr
                              \frac{L_2}{S}\right) + \\ &  + M_1\exp\left(-i\frac{\pi}{2}n\right)\exp\left(-i\frac{\pi}{4}\right)V \left(\kf
                              \frac{L_1+L_2}{S}-\kr\frac{L_2}{S}\right)~,
                          \end{split}
\end{equation} 
where 
\begin{equation}\label{index}
n=-2,-1,0,1~,
\end{equation}
\begin{equation}\label{m0}
M_0 = \frac{l h}{4},
\end{equation}
\begin{equation}\label{m1}
M_1 =  \frac{4}{\pi^3} l h  \sin\left(\frac{\pi}{4}\right),
\end{equation}
\begin{equation}
H_n^{(1,1)}  =
\frac{\sin\left(\pi k_2^f h\right)}{4\pi^3 k_1^f k_2^f}\left[\exp\left(i 2\pi k_1^f(n+1)l\right) - \exp\left(i 2\pi k_1^fnl\right)\right] \exp(i 2 \pi {\bf{k}}^f \cdot {\mathbf{t}})
\end{equation}

\begin{equation}
H_n^{(1,2)} = {\overline{H_n^{(1,1)}}}
\end{equation}

\begin{equation}
H_n^{(2,1)} =
 \frac{\sin\left(\pi k_2^r h\right)}{4\pi^3 k_1^r k_2^r}\left[\exp\left(i 2\pi k_1^r(n+1)l\right) - \exp\left(i 2\pi k_1^rn\right)\right] \exp(i 2 \pi {\bf{k}}^r \cdot {\mathbf{t}})
\end{equation}
\begin{equation}
H_n^{(2,2)} = {\overline{H_n^{(2,1)}}}
\end{equation}
\begin{equation}
\begin{split}
H_n^{(3,1)} = &
\frac{i \sin\left(\pi\left(k_2^f+k_2^r\right)h\right)}{2 \pi^4\left(k_1^f+k_1^r\right)\left(k_2^f+k_2^r\right)} \exp(i 2 \pi ({\bf{k}}^f+{\bf{k}}^r) \cdot {\mathbf{t}})
\cdot \\ & \cdot \left[\exp\left(i2\pi\left(k_1^f+k_1^r\right)(n+1)l\right) - \exp\left(i2\pi\left(k_1^f+k_1^r\right)nl\right)\right]
\end{split}
\end{equation}
\begin{equation}
H_n^{(3,2)}= {\overline{H_n^{(3,1)}}}~,
\end{equation}
and where the overline indicates the complex conjugate.
\end{theorem}
\proof We  denote with $P_n = [n l, (n+1)  l] \times [-h/2,h/2]$, $n=-2,-1,0,1$, the integration domain represented by the pixel and with ${\cal{S}}$ the integration domain on the solar disk (we point out that the values of $n$ are chosen in this way since the reference axes are centered in the middle of the detector as in Figure \ref{fig:detector}, right panel). Then 
\begin{equation}\label{counts-1}
{\cal{C}}_n = \int_{{\cal{S}}} \phi({\bf{x}}) \tau_n({\bf{x}}) d{\bf{x}},~~~n=-2,-1,0,1
\end{equation}
where 
\begin{equation}\label{tau}
\tau_n({\bf{x}}) = \int_{P_n} T(\x,\y)d{\bf{y}}, ~~~n=-2,-1,0,1
\end{equation}
and $T(\x,\y)$ is given in equation (\ref{product}). If just the first harmonic component is kept for each one of the two functions $F(\x,\y)$ and $R(\x,\y)$, i.e. 
\begin{equation}\label{front}
\begin{split}
F(\x,\y) \simeq & \frac{1}{2} + \frac{1}{ \pi i} \exp\left(i2\pi\kf\cdot\left(\y - \x\frac{L_1+L_2}{S} + {\mathbf{t}} \right)\right) - \\ & - \frac{1}{\pi i} \exp\left(-i2\pi\kf\cdot\left(\y - \x\frac{L_1+L_2}{S} + {\mathbf{t}} \right)\right)
\end{split}
\end{equation}
and 
\begin{equation}\label{rear}
\begin{split}
R(\x,\y)  \simeq & \frac{1}{2} + \frac{1}{ \pi i} \exp\left(i2\pi\kr\cdot\left(\y-\x\frac{L_2}{S} + {\mathbf{t}} \right)\right) - \\ & - \frac{1}{\pi i} \exp\left(-i2\pi\kr\cdot\left(\y-\x\frac{L_2}{S} + {\mathbf{t}} \right)\right)~,
\end{split}
\end{equation}
then the global transmission function becomes
\begin{equation}\label{transmission-2}
T({\bf{x}},{\bf{y}}) = \frac{1}{4} [T_1({\bf{x}},{\bf{y}}) + T_2({\bf{x}},{\bf{y}}) + T_3({\bf{x}},{\bf{y}}) + T_4({\bf{x}},{\bf{y}}) + T_5({\bf{x}},{\bf{y}}) + T_6({\bf{x}},{\bf{y}})],
\end{equation}
where
\begin{equation}\label{t1}
\begin{split}
T_1({\bf{x}},{\bf{y}}):= & 1 + \frac{2}{\pi i} \exp\left(i2\pi\kf\cdot\left(-\x\frac{L_1+L_2}{S}+\y + {\mathbf{t}} \right)\right) - \\ & - \frac{2}{\pi i} \exp\left(-i2\pi\kf\cdot\left(-\x
             \frac{L_1+L_2}{S}+\y + {\mathbf{t}} \right)\right)~,
\end{split}
\end{equation}
\begin{equation}\label{t2}
\begin{split}
T_2({\bf{x}},{\bf{y}}):= & \frac{2}{\pi i} \exp\left(i2\pi\kr\cdot\left(-\x\frac{L_2}{S}+\y {\mathbf{t}} \right)\right)- \\ & - \frac{2}{\pi i} \exp\left(-i2\pi\kr\cdot\left(-\x\frac{L_2}{S}+\y  +{\mathbf{t}} \right)\right)~,
\end{split}
\end{equation}
\begin{equation}\label{t3}
\begin{split}
T_3({\bf{x}},{\bf{y}}):= & - \frac{4}{\pi^2} \exp\left(-i2\pi\left(\frac{L_1+L_2}{S}\kf  +\frac{L_2}{S}\kr\right)\cdot \x \right) \cdot \\ & \cdot
\exp\left(i2\pi\left(\kf+\kr\right)\cdot \y \right) \exp\left(i2\pi\left(\kf+\kr\right)\cdot {\mathbf{t}} \right)~,
\end{split}
\end{equation} 
\begin{equation}\label{t4}
\begin{split}
T_4({\bf{x}},{\bf{y}}) := & -\frac{4}{\pi^2} \exp\left(i2\pi\left(\frac{L_1+L_2}{S}\kf+\frac{L_2}{S}\kr\right)\cdot\x\right) \cdot \\ & \cdot
\exp\left(-i2\pi\left(\kf+\kr\right)\cdot\y\right) \exp\left(-i2\pi\left(\kf+\kr\right)\cdot {\mathbf{t}} \right)~,
\end{split}
\end{equation}
\begin{equation}\label{t5}
\begin{split}
T_5({\bf{x}},{\bf{y}}):= & \frac{4}{\pi^2} \exp\left(-i2\pi\left(\frac{L_1+L_2}{S}\kf - \frac{L_2}{S}\kr\right)\cdot\x\right) \cdot \\ & \cdot
\exp\left(i2\pi\left(\kf-\kr\right)\cdot\y\right) \exp\left(i2\pi\left(\kf-\kr\right)\cdot {\mathbf{t}} \right)~,
\end{split}
\end{equation}
and
\begin{equation}\label{t6}
\begin{split}
T_6({\bf{x}},{\bf{y}}):= & \frac{4}{\pi^2} \exp\left(i2\pi\left(\frac{L_1+L_2}{S}\kf-\frac{L_2}{S}\kr\right)\cdot\x\right) \cdot \\ & \cdot
\exp\left(-i2\pi\left(\kf-\kr\right)\cdot\y\right) \exp\left(-i2\pi\left(\kf-\kr\right)\cdot {\mathbf{t}} \right).
\end{split}
\end{equation}
Including (\ref{transmission-2})-(\ref{t6}) into (\ref{tau}) leads to
\begin{equation}\label{tau-1}
\tau_n({\bf{x}}) = \psi_n^{(1)}({\bf{x}}) + \psi_n^{(2)}({\bf{x}}) + \psi_n^{(3)}({\bf{x}}) + \psi_n^{(4)}({\bf{x}}) + \psi_n^{(5)}({\bf{x}}) + \psi_n^{(6)}({\bf{x}}),
\end{equation}
where
\begin{equation}\label{tau1}
\begin{split}
\psi_n^{(1)}({\bf{x}}):= & M_0 + H_{n}^{(1,1)}\exp\left(-i2\pi\kf\frac{L_1+L_2}{S}\cdot\x\right) +  \\ & 
              + H_n^{(1,2)}\exp\left(i2\pi\kf\frac{L_1+L_2}{S}\cdot\x\right),
\end{split}
\end{equation}
\begin{equation}\label{tau2}
\psi_n^{(2)}({\bf{x}}) := H_n^{(2,1)}\exp\left(-i2\pi\kr\frac{L_2}{S}\cdot\x\right) + H_n^{(2,2)}\exp\left(i2\pi\kr\frac{L_2}{S}\cdot\x\right) ,
\end{equation}
\begin{equation}\label{tau3}
\psi_n^{(3)}({\bf{x}}):=H_n^{(3,1)}\exp\left(-i2\pi\left(\kf\frac{L_1+L_2}{S}+\kr\frac{L_2}{S}\right)\cdot\x\right) ,
\end{equation} 
\begin{equation}\label{tau4}
\psi_n^{(4)}({\bf{x}}) := H_n^{(3,2)}\exp \left(i2\pi\left(\kf\frac{L_1+L_2}{S}+\kr\frac{L_2}{S}\right)\cdot\x\right) ,
\end{equation}
\begin{equation}\label{tau5}
\psi_n^{(5)}({\bf{x}}):= M_1\exp\left(i\frac{\pi}{2}n\right)\exp\left(i\frac{\pi}{4}\right)\exp\left(-i2\pi\left(\kf\frac{L_1+L_2}{S}-\kr \frac{L_2}{S}\right)\cdot\x\right)~,
\end{equation}
and
\begin{equation}\label{tau6}
\psi_n^{(6)}({\bf{x}}):= M_1\exp\left(-i\frac{\pi}{2}n\right)\exp\left(-i\frac{\pi}{4}\right)\exp\left(i2\pi\left(\kf\frac{L_1+L_2}{S}-\kr  \frac{L_2}{S}\right)\cdot\x\right)~.
\end{equation}
Including (\ref{tau-1})-(\ref{tau6}) into (\ref{counts-1}) and using Defintion \ref{def-visibility} leads to (\ref{counts-3}).

\begin{remark}\label{rem-appr-coeff}
We notice that in equation (\ref{counts-3}) the terms containing the coefficients $H_{n}^{(1,1)}$, $H_n^{(1,2)}$, $H_n^{(2,1)}$, $H_n^{(2,2)}$, $H_n^{(3,1)}$ and $H_n^{(3,2)}$ are significantly smaller than the ones containing the coefficients $M_0$ and $M_1$. In fact, a direct numerical check shows that the former terms are smaller than the latter ones for orders of magnitude that range from $10^{-3}$ to $10^{-7}$. 
\end{remark}

\begin{figure}
\begin{center}
\includegraphics[width=9cm]{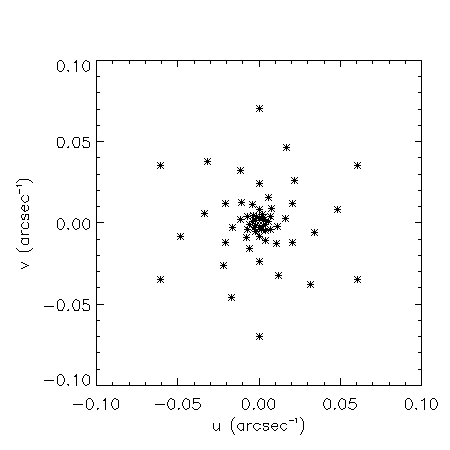}
\end{center}
\caption{Representation of the sampling of the spatial frequency plane in STIX where we have used the fact that the property $V(-\xik) = {\overline{V(\xik)}}$ allows the duplication of the $30$ spatial frequency samples.}
\label{fig:sampling}
\end{figure}

Theorem \ref{theo-counts} and Remark \ref{rem-appr-coeff} imply that an approximated estimate for the number of counts recorded by the $n-$th pixel in each {\em{STIX}} detector is
\begin{equation}\label{counts-4}
\begin{split}
{\cal{C}}_n \simeq & M_0V(\zero) + M_1\exp\left(i\frac{\pi}{2}n\right)\exp\left(i\frac{\pi}{4}\right)
                            V(-\xik) + \\ & + M_1\exp\left(-i\frac{\pi}{2}n\right)\exp\left(-i\frac{\pi}{4}\right)V(\xik)~,
 \end{split}
\end{equation}
with
\begin{equation}\label{csi}
\xik = {\bf{k}}^f \frac{L_1 + L_2}{S} - {\bf{k}}^r \frac{L_2}{S}~.
\end{equation}
Equation (\ref{csi}) indicates the positions in the spatial frequency plane where the sampling of the visibility is performed. Equation (\ref{csi}) depends through ${\bf{k}}^f$ and ${\bf{k}}^r$ on the geometrical properties, namely the orientation and the pitch size, of the grid pair (see equation [\ref{wavevector}]). Choosing a different collimator corresponds to choose a different set of values for the orientation and pitch size of the grids, which leads to a different $\xik$ point in the frequency plane. In this way the STIX grids' configuration draws the kind of sampling of the frequency plane represented in  Figure \ref{fig:sampling}.

As discussed in the previous Section, each {\em{STIX}} detector contains 4 pixels corresponding to the four values of $n=-2,-1,0,1$. It follows that the number of counts detected by the four pixels are
\begin{equation}\label{ca}
A:={\cal{C}}_{-2} \simeq M_0V({\bf{0}}) - M_1\exp\left(i\frac{\pi}{4}\right)V(-\xik)- M_1\exp\left(-i\frac{\pi}{4}\right)V(\xik) 
\end{equation}
\begin{equation}\label{cb}
B := {\cal{C}}_{-1} \simeq M_0V({\bf{0}}) -iM_1\exp\left(i\frac{\pi}{4}\right)V(-\xik)+iM_1\exp\left(-i\frac{\pi}{4}\right)V(\xik)
\end{equation}
\begin{equation}\label{cc}
C := {\cal{C}}_0 \simeq  M_0V({\bf{0}}) + M_1\exp\left(i\frac{\pi}{4}\right)V(-\xik)+M_1\exp\left(-i\frac{\pi}{4}\right)V(\xik)
\end{equation}
and
\begin{equation}\label{cd}
D:= {\cal{C}}_1 \simeq  M_0V({\bf{0}}) +iM_1\exp\left(i\frac{\pi}{4}\right)V(-\xik)-iM_1\exp\left(-i\frac{\pi}{4}\right)V(\xik)~,
\end{equation}
where, with a little abuse of notation, we indicated the counts recorded by each pixel with the same notations with which we have indicated the pixels in Figure \ref{fig:detector}, right panel. 
Therefore
\begin{equation}\label{cminusa}
C-A \simeq  4M_1{\rm{Re}} \left(\exp\left(-i\frac{\pi}{4}\right)V(\xik) \right)
\end{equation}
and
\begin{equation}\label{dminusb}
D-B \simeq 4M_1{\rm{Im}} \left(\exp\left(-i\frac{\pi}{4}\right)V(\xik) \right)~,
\end{equation}
where we have exploited the fact that $V(-\xik)={\overline{V(\xik)}}$, the overline representing the complex conjugation. 
Equations (\ref{cminusa}) and (\ref{dminusb}) imply that
\begin{equation}\label{visibilities-1}
V(\xik) \simeq \frac{1}{4M_1} \left[(C - A) +i (D - B)\right]\exp\left(i\frac{\pi}{4}\right)~.
\end{equation}
Please note that if the minus sign is taken in equation (\ref{difference-construction}), then the phase factor in equation (\ref{visibilities-1}) becomes $\exp(-i \pi/4)$.

%

\section{Approximation error}
Equations (\ref{cminusa}) and (\ref{dminusb}) connect the difference of the number of counts detected by the pixels in {\em{STIX}} collimators to the real and imaginary part of {\em{STIX}} visibilities, respectively. However, this interpretation is made possible by the first harmonic approximation of the Fourier series representing the transmission function for the front and rear grid of each collimator (see Theorem \ref{theo-counts} and Remark \ref{rem-appr-coeff}). In order to estimate the reliability of this approximation we consider the following numerical experiment. A point source is placed at the origin of the coordinate system in the Sun and the number of counts recorded by the $n$-th pixel in this physical configuration is analytically computed utilizing the complete Fourier series (\ref{bo-1}) and (\ref{bo-2}) for the two transmission functions (we point out that under these conditions, $\phi({\bf{x}})=\delta({\bf{x}})$ and $V(\xik) = 1$ for all sampled spatial frequencies). This requires the computation of
\begin{equation}\label{comput-1}
{\cal{C}}_n = \tau_n({\bf{0}})~~~~~n=-2,-1,0,1~,
\end{equation}
where
\begin{equation}\label{comput-2}
\tau_n({\bf{0}}) = \int_{P_n} T({\bf{0}},{\bf{y}})d{\bf{y}}~~~~~n=-2,-1,0,1~,
\end{equation}
and
\begin{equation}\label{comput-3}
T({\bf{0}},{\bf{y}}) = F({\bf{0}},{\bf{y}}) R({\bf{0}},{\bf{y}})~.
\end{equation}
Trigonometry and analytical integration over the pixel dimensions lead to 
\begin{equation}\label{comput-6}
{\cal{C}}_n = M_0 + I_1 + I_2 + I_3 + I_4
\end{equation}
where
\begin{equation}\label{comput-7}
I_1 =  \sum_{m=1,3}^{\infty} \frac{2\sin\left(\pi m k_1^fl\right)\sin\left(\pi m k_2^fh\right)}{\pi^3m^3 k^f_1 k^f_2}
 \sin\left(2\pi m\left(k_1^f\frac{2n+1}{2}l+k_2^f t\right)\right)
\end{equation}
\begin{equation}\label{comput-8}
I_2 =  \sum_{j=1,3}^{\infty} \frac{2\sin\left(\pi j k_1^rl\right)\sin\left(\pi j k_2^rh\right)}{\pi^3j^3 k^r_1 k^r_2}
 \sin\left(2\pi j\left(k_1^r\frac{2n+1}{2}l+k_2^rt\right)\right)
\end{equation}
\begin{equation}\label{comput-9}
\begin{split}
I_3  = & -\sum_{m=1,3}^{\infty}\sum_{j=1,3}^{\infty} \frac{2\sin\left(\pi l \left(mk_1^f+jk_1^r\right)\right)\sin\left(\pi h\left(mk_2^f+jk_2^r\right)\right)}{\pi^4mj(mk^f_1+jk^r_1)(mk^f_2+jk^r_2)} \cdot \\
& \cdot \cos\left(2\pi\left(\frac{2n+1}{2}l\left(mk_1^f+jk_1^r\right)-t\left(mk_2^f+jk_2^r\right)\right)\right)~.
\end{split}
\end{equation}
%
%
The computation of $I_4$ is complicated by the fact that $k_2^f - k_2^r=0$ and therefore the case $m=j$ must be determined separately. We obtained, for $m \neq j$:
\begin{equation}\label{comput-12}
I_4 = \sum_{m=1,3}^{\infty} \sum_{j=1,3}^{\infty} I_4^{(1)}(m,j) \cdot I_4^{(2)} (m,j)
\end{equation}
with
\begin{equation}\label{comput-13}
I_4^{(1)}(m,j) = \frac{2\sin\left(\pi l \left(mk_1^f-jk_1^r\right)\right)\sin\left(\pi h\left(mk_2^f-jk_2^r\right)\right)}{\pi^4mj(mk^f_1-jk^r_1)(mk^f_2-jk^r_2)}
\end{equation}
and
\begin{equation}\label{comput-14}
I_4^{(2)}(m,j) =  \cos\left(2\pi\left(\frac{2n+1}{2}l\left(mk_1^f-jk_1^r\right)-t\left(mk_2^f-jk_2^r\right)\right)\right)~;
\end{equation}
for $m=j$
\begin{equation}\label{comput-15}
I_4(m,m) = \frac{8hl}{\pi^3m^3}\sin\left(m\frac{\pi}{4}\right)\cos\left(\pi m\frac{2n+1}{4}\right)~.
\end{equation}
\begin{remark}\label{rem-appr-error}
Equations (\ref{comput-6})-(\ref{comput-15}) reduce to approximation (\ref{counts-4}) for $V(\xik) \equiv 1$, if $I_1$, $I_2$, and $I_3$ are neglected and just the first harmonic is kept in $I_4$, equation (\ref{comput-12}). We notice that $I_1$, $I_2$, and $I_3$ tend to zero at least as $m^{-3}$ where $m$ is the summation index in the series. On the other hand, we have that
\begin{equation}\label{I-H-1}
I_1(1) = H_{n}^{(1,1)} + H_{n}^{(1,2)}~,
\end{equation} 
\begin{equation}\label{I-H-2}
I_2(1) = H_{n}^{(2,1)} + H_{n}^{(2,2)}~,
\end{equation} 
and
\begin{equation}\label{I-H-3}
I_3(1,1) = H_{n}^{(3,1)} + H_{n}^{(3,2)}~,
\end{equation} 
where $I_1(1)$, $I_2(1)$, and $I_3(1,1)$ are the first terms in the series (\ref{comput-7}), (\ref{comput-8}), and (\ref{comput-9}), respectively. Therefore, as a consequence of Remark \ref{rem-appr-coeff}, we have that these terms are significantly smaller than $M_0$.
\end{remark}

\begin{figure}
\begin{center}
\begin{tabular}{cc}
\includegraphics[width=5.cm]{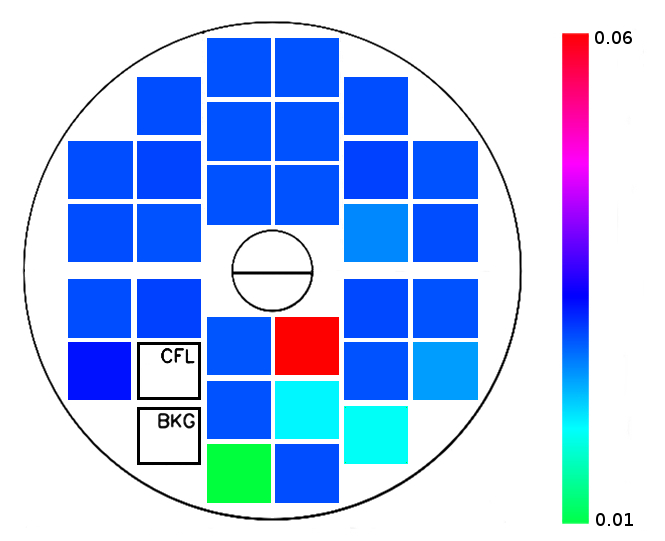} &
\includegraphics[width=5.cm]{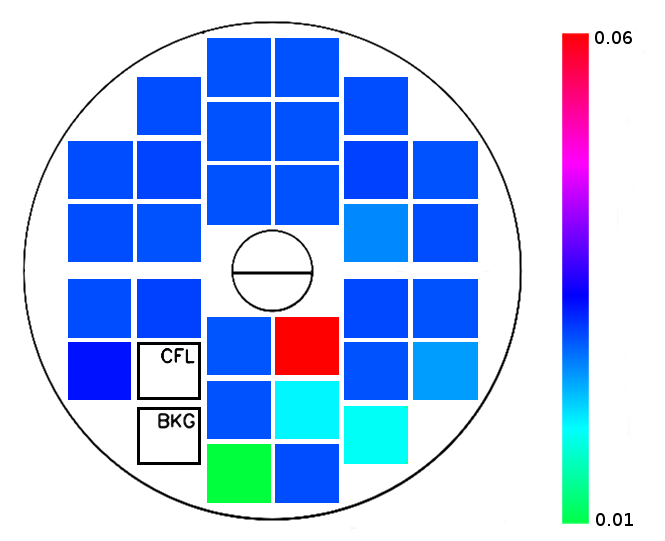} \\
\includegraphics[width=5.cm]{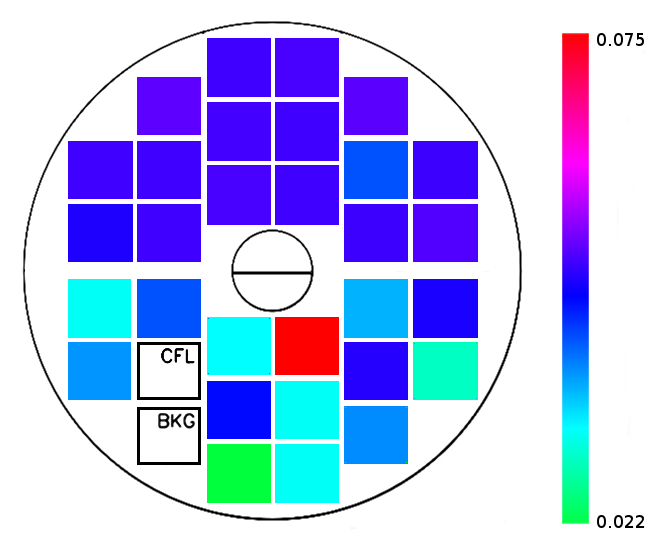} &
\includegraphics[width=5.cm]{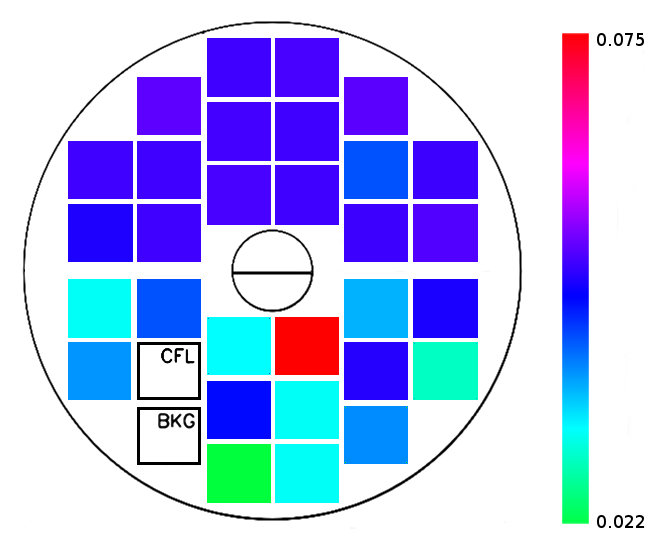}  
\end{tabular}
\end{center}
\caption{Pictorial representation of the relative approximation errors $\eta_1$ and $\eta_2$ in (\ref{relative-1}) and (\ref{relative-2}), respectively. This figure utilizes the actual layout of STIX, where each square corresponds to a detector and the color denotes the magnitude of the numerical errors (CFL stands for Coarse Flare Locator and BKG for Background monitor). Top left panel: values of $\eta_1$ when the source is a point source placed at ${\bf{x}}=0$ . Top right panel: values of $\eta_2$ when the source is a point source placed at ${\bf{x}}=0$. Bottom left panel: averaged values of $\eta_1$ with respect to $225$ positions of the point source, uniformly sampled in a square of side equal to $15$ arcsec centered at ${\bf{x}}=0$.. Bottom right panel: average values of $\eta_2$ with respect to $225$ positions of the point source, uniformly sampled in a square of side equal to $15$ arcsec centered at ${\bf{x}}=0$.}
\label{fig:error-visibilities}
\end{figure}

The availability of formulas (\ref{comput-6})-(\ref{comput-15}) allows the computation of the number of counts ${\cal{C}}_n$ recorded by pixel $n$ with whatever degree of accuracy, depending on the number of harmonics used in the formulas. 
Therefore we are now able to compute the relative error introduced by interpreting the real and imaginary parts of each visibility as differences of the number of counts recorded by the corresponding detector pixels as in equation (\ref{visibilities-1}). In fact, if 
\begin{equation}
\lambda_1 :=  4M_1 {\rm{Re}} \left(\exp\left(-i\frac{\pi}{4}\right)V(\xik) \right)~,
\end{equation}
as the right hand side of equation (\ref{cminusa}) and
\begin{equation}
\lambda_2 :=  4M_1 {\rm{Im}} \left(\exp\left(-i\frac{\pi}{4}\right)V(\xik) \right)~,
\end{equation}
as the right hand side of equation (\ref{dminusb}), we have the relative approximation errors
\begin{equation}\label{relative-1}
\eta_1 = \left| \frac{(C - A) - \lambda_1}{\lambda_1} \right|
\end{equation}
and
\begin{equation}\label{relative-2}
\eta_2 = \left| \frac{(D - B) - \lambda_2}{\lambda_2} \right|~,
\end{equation}
where $A,B,C,D$ can be computed by using a very large number of harmonic components (in the following experiment we will use $100$ components). In order to visually represent such errors, in Figure \ref{fig:error-visibilities}, top panels, we have utilized the actual layout of the $32$ {\em{STIX}} detector assembly to provide a colormap that quantifies such errors for each detector. We found that the relative errors $\eta_1$ and $\eta_2$ associated to approximation (\ref{visibilities-1}) always range between $1\%$ and $4\%$ except than in the case of sub-collimator $20$, where the relative error is around $6\%$. We notice that, coherently with this value of the error, sub-collimator $20$ is the one for which $I_1 + I_2 + I_3$ is maximum.

We analytically computed ${\cal{C}}_n$ in (\ref{comput-6}) again for $100$ harmonic components, but this time considering $225$ positions for the point source, uniformly sampled in a square of side equal to $15$ arcsec on the Sun, centered around ${\bf{x}}=0$. For each position and each sub-collimator we computed the approximation errors $\eta_1$ and $\eta_2$ as in (\ref{relative-1}) and (\ref{relative-2}); then, in Figure \ref{fig:error-visibilities}, bottom panels, for each sub-collimator we represented the average relative errors with respect to all positions. We found that these errors are in the range $2\%$ - $5\%$ for all sub-collimators except for sub-collimator $20$, for which the relative error is $7.5\%$.

\section{Conclusions}
We provided a mathematical model for signal formation in the {\em{Spectrometer/Telescope for Imaging X-rays (STIX)}} which will be part of the {\em{Solar Orbiter}} payload to be launched by {\em{ESA}} in either 2017 or 2018. Specifically, we computed the number of counts recorded by each {\em{STIX}} detector and showed the connection between this number and the concept of visibility in hard X-ray imaging. We also theoretically justified the way the spatial frequency plane is sampled by {\em{STIX}} sub-collimators and numerically discussed to what quantitative extent  first harmonic approximation is feasible in this context. 

The modeling of the {\em{STIX}} imaging concept in terms of visibilities has a valuable impact as far as the image reconstruction process is concerned. Several Fourier-based algorithms have been introduced for previous hard X-ray imaging missions \cite{boetal06,coetal85,ho74,maetal09,mapi13,pietal07,pretal09}, that utilize regularization to reduce artifacts provoked by under-sampling the $(u,v)$ plane. {\em{Ad hoc}} implementations of such algorithms for {\em{STIX}} visibilities are currently under construction.

\thanks{This work has been partly supported by an INdAM - GNCS Project 2014. We would also like to acknowledge Gordon Hurford, Federico Benvenuto and Anna Codispoti for useful discussions.}

\end{document}